\documentclass[a4paper,12pt]{article}
\usepackage{jheppubmod}
\usepackage{amsmath}
\usepackage{amssymb}
\usepackage{bbm}
\usepackage{amsfonts}
\usepackage{tipa}
\usepackage{mathrsfs}
\usepackage{mathtools}
\usepackage{graphicx}

\title{Higher Spin Theories from Finsler Geometry}

\author{Zhi-Qiang Guo}
\emailAdd{zhiqiang.guo@usm.cl}
\affiliation{Departamento de F\'{i}sica y Centro Cient\'{i}fico
Tecnol\'{o}gico de Valpara\'{i}so,\\ Universidad T\'{e}cnica Federico
Santa Mar\'{i}a,\\ Casilla 110-V, Valpara\'{i}so, Chile}

\abstract{We provide observations that Finsler geometry could be useful tools to construct higher-spin theories. We suggest that a Finsler metric of constant flag curvature can be regarded as a metric encoding higher-spin fields. We also show that the Fronsdal's equations for free higher-spin fields can be derived from equations of motion of constant curvature in Finsler geometry.}

\keywords{Higher-Spin Theory, Finsler Geometry}

\begin{document}
\maketitle

\section{Introduction}\label{sec:1}

The spin-1 gauge fields and the spin-2 gravity provide elegant descriptions of interactions of matter discovered so far. Despite of the lacking of experimental evidence, higher-spin fields also pose an important role in physics form theoretical perspectives. Our understanding on higher-spin theories has made steady progress in the past several decades~\cite{Bekaert:2005vh,Bekaert:2010hw}. Along with the frame-like approach or gauge approach, Vasiliev has proposed non-linear equations of motion for higher-spin fields~\cite{Vasiliev:1990en,Vasiliev:2003ev}. These non-linear equations were used to show the consistency~\cite{Giombi:2009wh} about the conjectured correspondence between the higher-spin theory and the $O(N)$ vector model~\cite{Sezgin:2001zs,Klebanov:2002ja}.  Consistent cubic interaction vertices have been constructed~\cite{Fradkin:1986qy,Boulanger:2012dx}, albeit a complete action for Vasiliev theory is still under explorations~\cite{Boulanger:2011dd,Doroud:2011xs}.

Much earlier than the frame-like approach,  the metric-like approach or geometrical approach has been pursued along several different lines. For free massless higher-spin fields, Fronsdal proposed second derivative actions with restricted field variables and gauge transformations~\cite{Fronsdal:1978rb,Fang:1978wz}. de Wit and Freedman found that such kind of constraints on field variables can be understood naturally by considering generalized Christoffel symbols with higher derivatives~\cite{deWit:1979pe}, which are further used to construct equations of motions for non-restricted field variables albeit with non-local formulations~\cite{Francia:2002aa,Francia:2002pt}. On the level of Lagrangian, consistent cubic interaction vertices were also constructed from the Noether procedure~\cite{Manvelyan:2010wp,Manvelyan:2010jf,Joung:2011ww,Joung:2012fv} and the BRST procedure~\cite{Fotopoulos:2008ka}. For the metric-like approach, it seems difficult to construct actions for higher-spin fields beyond cubic orders and also difficult to formulate appropriate equations of motion because of the lacking of general guiding principles.

In this paper, we attempt to propose that Finsler geometry~\cite{bao2000,chern2005,miron1994,miron2010} could be useful tools to construct theories for higher-spin fields. The feasibility of dealing with higher-spin fields in the framework of Finsler geometry can be traced back to an observation of de Wit and Freedman. Instead of constructing actions for interactive higher-spin fields, de Wit and Freedman~\cite{deWit:1979pe} proposed an action which describes interactions among particles and higher-spin fields
\begin{eqnarray}
\label{sec1-dewit-action}
I=\int_{-\infty}^{\infty}dt\left[\left(\eta_{\mu\nu}\frac{dx^{\mu}}{dt}\frac{dx^{\nu}}{dt}\right)^{\frac{1}{2}}
+e\left(\eta_{\mu\nu}\frac{dx^{\mu}}{dt}\frac{dx^{\nu}}{dt}\right)^{-\frac{3}{2}}\varphi_{\alpha\beta\rho\sigma}
\frac{dx^{\alpha}}{dt}\frac{dx^{\beta}}{dt}\frac{dx^{\rho}}{dt}\frac{dx^{\sigma}}{dt}\right].
\end{eqnarray}
This action includes the conventional spin-2 field and also higher-spin fields. Besides the invariance under the general coordinate transformations, de Wit and Freedman observed that the action~(\ref{sec1-dewit-action}) is also invariant up to second order of the coupling constant $e$ under the transformations
\begin{eqnarray}
\label{sec1-dewit-trans-1}
\delta \varphi_{\alpha\beta\rho\sigma}&=&4\partial_{(\alpha}\xi_{\beta\rho\sigma)},\\
\label{sec1-dewit-trans-2}
\delta{x^{\alpha}}&=&-12e\left(\eta_{\mu\nu}\frac{dx^{\mu}}{dt}\frac{dx^{\nu}}{dt}\right)^{-\frac{3}{2}}
\xi^{\alpha}_{\hspace{1mm}\beta\rho\sigma}
\frac{dx^{\beta}}{dt}\frac{dx^{\rho}}{dt}\frac{dx^{\sigma}}{dt}.
\end{eqnarray}
The transformations~(\ref{sec1-dewit-trans-1}) is just the expected Abelian gauge transformation for higher-spin fields. Another point we can realize is that the action~(\ref{sec1-dewit-action}) can be regarded as the definition of the line element in Finsler geometry. The above two observations make us conjecture that some special geometrical quantities constructed from Finsler geometry could be appropriate objects to describe higher-spin theories. Of course, we are confronted with problems immediately at this moment. As is known, Finsler geometry is mainly used as tools to describe the anisotropic space-time~\cite{Girelli:2006fw,Bogoslovsky:2007gt,Gibbons:2007iu,Chang:2008yv,antonelli1993,Vacaru:2008qs}, that is, the metric in Finsler geometry depends on unspecified tangent vectors. Hence we need overcome this barrier and construct geometrical quantities which are independent of these tangent vectors. We shall show that this mentioned problem can be overcome by considering equations of motion of constant curvature appropriately.

The left of this paper is organized as follows. In section \ref{sec:2}, we present a simple introduction to Finsler geometry, where the necessary quantities for our paper are defined. In section \ref{sec:3.1}, we propose interpretations that the Funk metric of constant curvature can be regarded as a metric encoding higher-spin fields. In section \ref{sec:3.2}, we derive Fronsdal's equations from the equation of constant curvature taking the spin-4 field as an example. The possible non-linear extensions are discussed in section \ref{sec:3.3}. Finally, we give some discussions and conclusions in section \ref{sec:4}.

\section{Basics of Finsler Geometry}\label{sec:2}

As a generalization of Riemannian geometry, Finsler geometry has been mentioned by Riemann in his 1854 lecture~\cite{riemann1953}. In Riemannian geometry, the line-element is defined by
\begin{eqnarray}
\label{sec2-line-rie}
\frac{ds}{dt}=\sqrt{\tilde{g}_{\mu\nu}(x)\frac{dx^{\mu}}{dt}\frac{dx^{\nu}}{dt}}.
\end{eqnarray}
Instead of restrictions on square dependence on $\frac{dx}{dt}$, we can consider a more general definition of the line-element
\begin{eqnarray}
\label{sec2-line-fin}
\frac{ds}{dt}=F(x,\frac{dx}{dt}).
\end{eqnarray}
Here the Finsler function $F$ is a homogeneous function of $\frac{dx}{dt}$, which satisfies
\begin{eqnarray}
\label{sec2-line-fin-homo}
F(x,\lambda\frac{dx}{dt})=\lambda {F}(x,\frac{dx}{dt}).
\end{eqnarray}
The metric in Finsler geometry can be determined from this Finsler function by
\begin{eqnarray}
\label{sec2-line-fin-metr}
g_{\mu\nu}(x,y)=\frac{1}{2}\frac{\partial^2\tilde{F}}{\partial y^{\mu}\partial y^{\nu}}.
\end{eqnarray}
Here we has used the symbol $y^{\mu}=\frac{dx^{\mu}}{dt}$ and $\tilde{F}=F^2$ for convenience. Two important objects in Finsler geometry are the Cartan tensor
\begin{eqnarray}
\label{sec2-line-fin-cartan}
C_{\alpha\mu\nu}(x,y)=\frac{1}{4}\frac{\partial^3\tilde{F}}{\partial y^{\alpha}\partial y^{\mu}\partial y^{\nu}}=\frac{1}{2}\frac{\partial {g}_{\mu\nu}}{\partial y^{\alpha}}
\end{eqnarray}
and the non-linear connection
\begin{eqnarray}
\label{sec2-line-fin-nlc}
N^{\alpha}_{\hspace{1mm}\beta}&=&\gamma^{\alpha}_{\hspace{1mm}\beta\mu}y^{\mu}
-C^{\alpha}_{\hspace{1mm}\beta\mu}\gamma^{\mu}_{\hspace{1mm}\rho\sigma}y^{\rho}y^{\sigma},\\
\label{sec2-line-fin-nlc-ch}
\gamma^{\alpha}_{\hspace{1mm}\beta\mu}&=&g^{\alpha\nu}\gamma_{\nu\beta\mu},
\hspace{2mm}\gamma_{\nu\beta\mu}=\frac{1}{2}\left(\frac{\partial {g}_{\beta\nu}}{\partial {x}^{\mu}}+\frac{\partial {g}_{\mu\nu}}{\partial {x}^{\beta}}-\frac{\partial {g}_{\beta\mu}}{\partial {x}^{\nu}}\right).
\end{eqnarray}
For definitions of covariant derivatives, we need associated connections which behave like the Levi-Civita connection in Riemannian geometry. There are several associated connections in Finsler geometry, such as, the Cartan connection, the Berwald connection, the Chern-Rund connection and the Himaguchi connection~\cite{bao2000}. In this paper, we use the Chern-Rund connection for simplicity. The Chern-Rund connection is defined by
\begin{eqnarray}
\label{sec2-line-fin-crc}
\Gamma^{\alpha}_{\hspace{1mm}\beta\mu}&=&\frac{1}{2}g^{\alpha\nu}\left(\frac{\delta {g}_{\beta\nu}}{\delta {x}^{\mu}}+\frac{\delta {g}_{\mu\nu}}{\delta {x}^{\beta}}-\frac{\delta {g}_{\beta\mu}}{\delta {x}^{\nu}}\right),\hspace{3mm}\frac{\delta}{\delta {x}^{\mu}}=\frac{\partial}{\partial {x}^{\mu}}-N^{\alpha}_{\hspace{1mm}\mu}\frac{\partial}{\partial {y}^{\alpha}},\nonumber\\
&=&\gamma^{\alpha}_{\hspace{1mm}\beta\mu}-
g^{\alpha\rho}(C_{\rho\beta\sigma}N^{\sigma}_{\hspace{1mm}\mu}+C_{\rho\mu\sigma}N^{\sigma}_{\hspace{1mm}\beta}
-C_{\beta\mu\sigma}N^{\sigma}_{\hspace{1mm}\rho}).
\end{eqnarray}
The Chern-Rund connection is a torsion-free connection, as it is symmetrical about its indices $\beta$ and $\mu$. Utilizing the Chern-Rund connection, the horizontal covariant derivative for a tensor $T^{\alpha}_{\hspace{1mm}\beta}$ can be given by
\begin{eqnarray}
\label{sec2-line-fin-cd}
\nabla_{\mu}T^{\alpha}_{\hspace{1mm}\beta}=\frac{\delta}{\delta {x}^{\mu}}T^{\alpha}_{\hspace{1mm}\beta}
-\Gamma^{\rho}_{\hspace{1mm}\beta\mu}T^{\alpha}_{\hspace{1mm}\rho}
+\Gamma^{\alpha}_{\hspace{1mm}\mu\rho}T^{\rho}_{\hspace{1mm}\beta}.
\end{eqnarray}
The Chern-Rund connection is also compatible with the horizontal covariant derivative
\begin{eqnarray}
\label{sec2-line-fin-cd-compa}
\nabla_{\mu}g_{\alpha\beta}=\frac{\delta}{\delta {x}^{\mu}}g_{\alpha\beta}
-\Gamma^{\rho}_{\hspace{1mm}\beta\mu}g_{\alpha\rho}
-\Gamma^{\rho}_{\hspace{1mm}\alpha\mu}g_{\rho\beta}=0.
\end{eqnarray}
The horizontal Curvature in Finsler geometry can be derived from the commutator of horizontal covariant derivatives
\begin{eqnarray}
\label{sec2-line-fin-curvature}
[\nabla_{\mu},\nabla_{\nu}]V^{\sigma}&=&R^{\hspace{1mm}\sigma}_{\rho\hspace{1mm}\mu\nu}V^{\rho},\\
R^{\hspace{1mm}\sigma}_{\rho\hspace{1mm}\mu\nu}&=&\frac{\delta}{\delta {x}^{\mu}}\Gamma^{\sigma}_{\hspace{1mm}\rho\nu}-\frac{\delta}{\delta {x}^{\nu}}\Gamma^{\sigma}_{\hspace{1mm}\rho\mu}+\Gamma^{\sigma}_{\hspace{1mm}\alpha\mu}\Gamma^{\alpha}_{\hspace{1mm}\rho\nu}
-\Gamma^{\sigma}_{\hspace{1mm}\alpha\nu}\Gamma^{\alpha}_{\hspace{1mm}\rho\mu}.
\end{eqnarray}

Instead of using the metric-like field as above, we can also use the frame-like field in Finsler geometry. For the Finsler metric in (\ref{sec2-line-fin-metr}), we can construct a group of orthogonal basis $e_{a}^{\mu}$ satisfying
\begin{eqnarray}
\label{sec2-line-fin-frame}
e_{a}^{\mu}e_{b}^{\nu} g_{\mu\nu}={\eta}_{ab}
\end{eqnarray}
and the inverse frame basis $\theta_{\mu}^{a}$ satisfying
\begin{eqnarray}
\label{sec2-line-fin-frame-inverse}
e_{a}^{\mu}\theta_{\mu}^{b} ={\delta}_{a}^{b},\hspace{2mm} e_{a}^{\mu}\theta_{\nu}^{a} ={\delta}_{\nu}^{\mu},\hspace{1mm}g_{\mu\nu}={\eta}_{ab}\theta_{\mu}^{a}\theta_{\nu}^{b}.
\end{eqnarray}
As in Riemannian geometry, the spin connection $\omega_{b}^{\hspace{1mm}a}$ can be expressed by the frame-like field through the torsion-free condition. For the Chern-Rund connection (\ref{sec2-line-fin-metr}), the spin connection is given by
\begin{eqnarray}
\label{sec2-line-fin-spin-con}
\omega_{b}^{\hspace{1mm}a}=\theta_{\mu}^{a}de_{b}^{\mu}
+e_{b}^{\beta}\theta_{\alpha}^{a}\Gamma^{\alpha}_{\hspace{1mm}\beta\mu}dx^{\mu}.
\end{eqnarray}
We notice that $e_{a}^{\mu}$ depends on $(x,y)$, so the exterior differentiation for a function $f(x,y)$ acts as $df=\frac{\partial {f}}{\partial {x}^{\mu}}dx^{\mu}+\frac{\partial {f}}{\partial {y}^{\mu}}dy^{\mu}$. The curvature can be derived from
\begin{eqnarray}
\label{sec2-line-fin-spin-cur}
\Omega_{b}^{\hspace{1mm}a}&=&d\omega_{b}^{\hspace{1mm}a}+\omega_{c}^{\hspace{1mm}a}\wedge \omega_{b}^{\hspace{1mm}c}\\
\label{sec2-line-fin-spin-cur-1}
&=&R^{\hspace{1mm}a}_{b\hspace{1mm}cd}\theta^{c}\wedge \theta^{d}+P^{\hspace{1mm}a}_{b\hspace{1mm}cd}\theta^{c}\wedge \theta^{n+d},
\end{eqnarray}
where $\theta^{a}=\theta_{\mu}^{a}dx^{\mu}$ and $\theta^{n+a}=\theta_{\mu}^{a}\delta {y}^{\mu}$ are the 1-form field; while $\delta {y}^{\mu}=d{y}^{\mu}+N^{\mu}_{\hspace{1mm}\alpha}d{x}^{\alpha}$ is the dual basis for $\frac{\partial}{\partial{y}^{\mu}}$. The curvature $R^{\hspace{1mm}a}_{b\hspace{1mm}cd}$ in (\ref{sec2-line-fin-spin-cur-1}) is connected to the curvature $R^{\hspace{1mm}\sigma}_{\rho\hspace{1mm}\mu\nu}$ in (\ref{sec2-line-fin-curvature}) by
\begin{eqnarray}
\label{sec2-line-fin-spin-cur-link}
R^{\hspace{1mm}a}_{b\hspace{1mm}cd}=\theta_{\sigma}^{a}e_{b}^{\rho}e_{c}^{\mu}e_{d}^{\nu}
R^{\hspace{1mm}\sigma}_{\rho\hspace{1mm}\mu\nu}.
\end{eqnarray}

\section{Equations of Motion for Higher-Spin Fields}\label{sec:3}

\subsection{A Higher-Spin Metric of Constant Flag Curvature}\label{sec:3.1}

The Finsler geometry is widely used as tools to deal with anisotropic space-time~\cite{Girelli:2006fw,Bogoslovsky:2007gt,Gibbons:2007iu,Chang:2008yv,antonelli1993,Vacaru:2008qs}, because the metric (\ref{sec2-line-fin-metr}) in Finsler geometry depends on an unspecified vector generally. However, in this section, we attempt to argue that the Finsler geometry actually can also be regarded as theories to describe higher-spin fields, as can be understood from the following example. Contracting $y^{\mu}$ twice with the curvature $R^{\hspace{1mm}\sigma}_{\rho\hspace{1mm}\mu\nu}$, we can obtain a second order tensor
\begin{eqnarray}
\label{sec3-cur-2-order}
\mathcal{R}^{\alpha}_{\hspace{1mm}\beta}&=&y^{\mu}y^{\nu}R^{\hspace{1mm}\alpha}_{\mu\hspace{1mm}\beta\nu}
=2\frac{\partial {G}^{\alpha}}{\partial {x}^{\beta}}-{y}^{\mu}\frac{\partial^2 {G}^{\alpha}}{\partial {x}^{\mu}\partial{y}^{\beta}}+2{G}^{\mu}\frac{\partial^2 {G}^{\alpha}}{\partial {y}^{\mu}\partial{y}^{\beta}}-\frac{\partial{G}^{\alpha}}{\partial {y}^{\mu}}\frac{\partial {G}^{\mu}}{\partial{y}^{\beta}},\\
\label{sec3-cur-coeff-spray}
{G}^{\mu}&=&\frac{1}{4}g^{\mu\lambda}\left({y}^{\nu}\frac{\partial^2 \tilde{F}}{\partial {x}^{\nu}{y}^{\lambda}}-\frac{\partial\tilde{F}}{\partial {x}^{\lambda}}\right)=\frac{1}{2}\gamma^{\mu}_{\hspace{1mm}\rho\sigma}{y}^{\rho}{y}^{\sigma}.
\end{eqnarray}
Here $G^{\mu}$ are the spray coefficients, which are another important quantities in Finsler geometry~\cite{chern2005}. The 2nd class of Christoffel symbols are defined by Eq.~(\ref{sec2-line-fin-nlc-ch}). A Finsler metric with the constant flag curvature satisfies the equation
\begin{eqnarray}
\label{sec3-cur-con-flag}
\mathcal{R}^{\alpha}_{\hspace{1mm}\beta}=\Lambda(F^2\delta^{\alpha}_{\hspace{1mm}\beta}-g_{\beta\mu}y^{\mu}y^{\alpha}),
\end{eqnarray}
or in another way
\begin{eqnarray}
\label{sec3-cur-con-flag-scalar}
\mathcal{R}=\mathcal{R}^{\alpha}_{\hspace{1mm}\alpha}=\Lambda(d-1)F^2.
\end{eqnarray}
Here $d$ is the dimension of space-time. We consider the Funk metric\footnote{For other metrics of constant flag curvature, see the chapter ``Landsberg Curvature, S-Curvature and Riemann Curvature" written by Z. Shen in the book~\cite{Bao2004}.}
\begin{eqnarray}
\label{sec3-funk-met}
F=\frac{\sqrt{(1-\omega^2{x}_{\mu}{x}^{\mu}){y}_{\mu}{y}^{\mu}+\omega^2({x}_{\mu}{y}^{\mu})^2}}{1-\omega^2{x}_{\mu}{x}^{\mu}}
+\frac{\omega{x}_{\mu}{y}^{\mu}}{1-\omega^2{x}_{\mu}{x}^{\mu}},
\end{eqnarray}
where ${x}_{\mu}{x}^{\mu}=\eta_{\mu\nu}{x}^{\mu}{x}^{\nu}$. The Funk metric satisfies the Hamel equation
\begin{eqnarray}
\label{sec3-funk-hamel}
\frac{\partial}{\partial {y}^{\lambda}}\left[{y}^{\nu}\frac{\partial{F}}{\partial {x}^{\nu}}\right]=2\frac{\partial{F}}{\partial {x}^{\lambda}}.
\end{eqnarray}
It is projectively flat and through calculations we have
\begin{eqnarray}
\label{sec3-funk-hamel-pro-flat}
{G}^{\mu}=P(x,y){y}^{\mu},\hspace{2mm}P=\frac{\omega}{2}F,\hspace{2mm}\mathcal{R}=-\frac{\omega^2}{4}(d-1)F^2.
\end{eqnarray}
So the Funk metric has the constant flag curvature $-\frac{\omega^2}{4}$. The Funk metric is of the Randers type~\cite{Randers:61}
\begin{eqnarray}
\label{sec3-funk-randers}
F&=&\sqrt{a_{\mu\nu}{y}^{\mu}{y}^{\nu}}+b_{\mu}{y}^{\mu},\\
\label{sec3-funk-randers-1}
a_{\mu\nu}&=&
\frac{(1-\omega^2{x}_{\lambda}{x}^{\lambda})\eta_{\mu\nu}+\omega^2{x}_{\mu}{x}_{\nu}}{(1-\omega^2{x}_{\lambda}{x}^{\lambda})^2},
\hspace{2mm}
b_{\mu}=\frac{\omega{x}_{\mu}}{1-\omega^2{x}_{\lambda}{x}^{\lambda}}.
\end{eqnarray}
To obtain the metric tensor through Eq.~(\ref{sec2-line-fin-metr}), we need the square of the Finsler function
\begin{eqnarray}
\label{sec3-funk-square}
\tilde{F}=F^2=a+b^2+2b\sqrt{a},\hspace{2mm}a=a_{\mu\nu}{y}^{\mu}{y}^{\nu},\hspace{2mm}b=b_{\mu}{y}^{\mu},
\end{eqnarray}
which can be expanded as
\begin{eqnarray}
\label{sec3-funk-square-series}
\tilde{F}&=&a+b^2+\sqrt{(a+b^2)^2-(a-b^2)^2}\nonumber\\
&=&a+b^2+(a+b^2)\left[1-\frac{1}{2}\frac{(a-b^2)^2}{(a+b^2)^2}-\frac{1}{8}\frac{(a-b^2)^4}{(a+b^2)^4}+\cdots\right]\nonumber\\
&=&2(a+b^2)-\frac{1}{2}\frac{(a-b^2)^2}{a+b^2}-\frac{1}{8}\frac{(a-b^2)^4}{(a+b^2)^3}+\cdots.
\end{eqnarray}
We see that it has the pattern of the de Wit-Freedman metric
\begin{eqnarray}
\label{sec3-funk-square-series-df}
\tilde{F}=\tilde{a}_{\mu\nu}{y}^{\mu}{y}^{\nu}
-\frac{A_{\alpha\beta\mu\nu}{y}^{\alpha}{y}^{\beta}{y}^{\mu}{y}^{\nu}}{\tilde{a}_{\mu\nu}{y}^{\mu}{y}^{\nu}}
-\frac{A_{\alpha\beta\mu\nu\rho\sigma\lambda\tau}{y}^{\alpha}{y}^{\beta}{y}^{\mu}{y}^{\nu}{y}^{\rho}{y}^{\sigma}{y}^{\lambda}{y}^{\tau}}
{(\tilde{a}_{\mu\nu}{y}^{\mu}{y}^{\nu})^3}+\cdots,
\end{eqnarray}
with the identifications
\begin{eqnarray}
\label{sec3-funk-square-series-df-id-2}
\tilde{a}_{\mu\nu}
&=&2\frac{(1-\omega^2{x}_{\lambda}{x}^{\lambda})\eta_{\mu\nu}+2\omega^2{x}_{\mu}{x}_{\nu}}{(1-\omega^2{x}_{\lambda}{x}^{\lambda})^2},\\
\label{sec3-funk-square-series-df-id-4}
A_{\alpha\beta\mu\nu}&=&\frac{1}{3}\frac{1}{(1-\omega^2{x}_{\lambda}{x}^{\lambda})^2}
(\eta_{\alpha\beta}\eta_{\mu\nu}+\eta_{\alpha\mu}\eta_{\beta\nu}+\eta_{\alpha\nu}\eta_{\mu\beta}).
\end{eqnarray}
Here the fields $A_{\alpha\beta\mu\nu}$ et al. are totally symmetrical about their indices. So we may regard the Funk metric as a metric encoding the conventional spin-2 field and more higher-spin fields. In other way, supposing that we have a group of higher-spin fields taking values as in Eqs.~(\ref{sec3-funk-square-series-df-id-2})-(\ref{sec3-funk-square-series-df-id-4}), then they can define a Funk metric through Eq.~(\ref{sec3-funk-square-series-df}) which has the constant flag curvature. This observation suggests that the Finsler geometry could be regraded as theories dictating interactions of higher-spin fields.

\subsection{Equations of Motion}\label{sec:3.2}

A next important point is whether we can derive equations of motion of higer-spin fields from Finsler geometry. For this purpose, we begin with a more general Finsler function as
\begin{eqnarray}
\label{sec3-finsler-def}
\tilde{F}=F^2=A^{(2)}+\frac{A^{(4)}}{A^{(2)}}+\frac{A^{(6)}}{[A^{(2)}]^{2}}+\sum_{s=8}^{\infty}\frac{A^{(s)}}{[A^{(2)}]^{\frac{s-2}{2}}},
\end{eqnarray}
with the definition
\begin{eqnarray}
\label{sec3-finsler-def-s}
A^{(s)}=A_{\mu_1\mu_2\cdots\mu_s}(x)y^{\mu_1}y^{\mu_2}\cdots{y}^{\mu_s}.
\end{eqnarray}
Here we only consider fields with even indices. The fields with odd indices introduce irrational functions like the root of square, which might cause problems for our subsequent applications. Through Eq.~(\ref{sec2-line-fin-metr}), we obtain the Finsler metric as
\begin{eqnarray}
\label{sec3-finsler-def-met}
g_{\alpha\beta}=A_{\alpha\beta}+\frac{1}{A^{(2)}}g^{(1)}_{\alpha\beta}+\frac{1}{[A^{(2)}]^{2}}g^{(2)}_{\alpha\beta}
+\frac{1}{[A^{(2)}]^{3}}g^{(3)}_{\alpha\beta}+\cdots,
\end{eqnarray}
with the identifications
\begin{eqnarray}
\label{sec3-finsler-def-met-1}
g^{(1)}_{\alpha\beta}&=&6A^{(4)}_{\alpha\beta},\\
\label{sec3-finsler-def-met-2}
g^{(2)}_{\alpha\beta}&=&15A^{(6)}_{\alpha\beta}-A^{(2)}_{\alpha\beta}A^{(4)}-4(A^{(2)}_{\alpha}A^{(4)}_{\beta}
+A^{(4)}_{\alpha}A^{(2)}_{\beta}),\\
\label{sec3-finsler-def-met-3}
g^{(3)}_{\alpha\beta}&=&28A^{(8)}_{\alpha\beta}-2A^{(2)}_{\alpha\beta}A^{(6)}-12(A^{(2)}_{\alpha}A^{(6)}_{\beta}
+A^{(6)}_{\alpha}A^{(2)}_{\beta})+2A^{(2)}_{\alpha}A^{(2)}_{\beta}A^{(4)},
\end{eqnarray}
and here we have used the symbols
\begin{eqnarray}
\label{sec3-finsler-def-met-sder}
A^{(s)}_{\alpha}=A_{\alpha\mu_1\mu_2\cdots\mu_{s-1}}(x)y^{\mu_1}y^{\mu_2}\cdots{y}^{\mu_{s-1}},\hspace{2mm}
A^{(s)}_{\alpha\beta}=A_{\alpha\beta\mu_1\mu_2\cdots\mu_{s-2}}(x)y^{\mu_1}y^{\mu_2}\cdots{y}^{\mu_{s-2}},\hspace{2mm}\cdots.
\end{eqnarray}
Through Eq.~(\ref{sec2-line-fin-nlc-ch}), the 1st class of Christoffel symbol $\gamma_{\lambda\alpha\beta}$ has the same pattern with $g_{\alpha\beta}$
\begin{eqnarray}
\label{sec3-cur-coeff-spray-series-2ch}
\gamma_{\lambda\alpha\beta}=\gamma^{(0)}_{\lambda\alpha\beta}+\frac{1}{A^{(2)}}\gamma^{(1)}_{\lambda\alpha\beta}
+\frac{1}{[A^{(2)}]^{2}}\gamma^{(2)}_{\lambda\alpha\beta}
+\frac{1}{[A^{(2)}]^{3}}\gamma^{(3)}_{\lambda\alpha\beta}+\cdots,
\end{eqnarray}
with
\begin{eqnarray}
\label{sec3-cur-coeff-spray-series-0-2ch}
\gamma^{(0)}_{\lambda\alpha\beta}&=&\frac{1}{2}\left[\frac{\partial{A}_{\lambda\beta}}{\partial{x}^{\alpha}}
+\frac{\partial{A}_{\lambda\alpha}}{\partial{x}^{\beta}}-\frac{\partial{A}_{\alpha\beta}}{\partial{x}^{\lambda}}\right],\\
\label{sec3-cur-coeff-spray-series-1-2ch}
\gamma^{(1)}_{\lambda\alpha\beta}&=&\frac{1}{2}\left[\frac{\partial{g}^{(1)}_{\lambda\beta}}{\partial{x}^{\alpha}}
+\frac{\partial{g}^{(1)}_{\lambda\alpha}}{\partial{x}^{\beta}}-\frac{\partial{g}^{(1)}_{\alpha\beta}}{\partial{x}^{\lambda}}\right],
\end{eqnarray}
and
\begin{eqnarray}
\label{sec3-cur-coeff-spray-series-2-2ch}
\gamma^{(2)}_{\lambda\alpha\beta}&=&-\frac{1}{2}\left[{g}^{(1)}_{\lambda\beta}\frac{\partial{A}^{(2)}}{\partial{x}^{\alpha}}
+{g}^{(1)}_{\lambda\alpha}\frac{\partial{A}^{(2)}}{\partial{x}^{\beta}}-{g}^{(1)}_{\alpha\beta}\frac{\partial{A}^{(2)}}{\partial{x}^{\lambda}}\right]
+\frac{1}{2}\left[\frac{\partial{g}^{(2)}_{\lambda\beta}}{\partial{x}^{\alpha}}
+\frac{\partial{g}^{(2)}_{\lambda\alpha}}{\partial{x}^{\beta}}-\frac{\partial{g}^{(2)}_{\alpha\beta}}{\partial{x}^{\lambda}}\right],\\
\label{sec3-cur-coeff-spray-series-3-2ch}
\gamma^{(3)}_{\lambda\alpha\beta}&=&-\frac{1}{2}\left[{g}^{(2)}_{\lambda\beta}\frac{\partial{A}^{(2)}}{\partial{x}^{\alpha}}
+{g}^{(2)}_{\lambda\alpha}\frac{\partial{A}^{(2)}}{\partial{x}^{\beta}}-{g}^{(2)}_{\alpha\beta}\frac{\partial{A}^{(2)}}{\partial{x}^{\lambda}}\right]
+\frac{1}{2}\left[\frac{\partial{g}^{(3)}_{\lambda\beta}}{\partial{x}^{\alpha}}
+\frac{\partial{g}^{(3)}_{\lambda\alpha}}{\partial{x}^{\beta}}-\frac{\partial{g}^{(3)}_{\alpha\beta}}{\partial{x}^{\lambda}}\right].
\end{eqnarray}
For convenience which will be clear later on, we rewrite Eqs.~(\ref{sec3-cur-coeff-spray-series-2-2ch}) as
\begin{eqnarray}
\label{sec3-cur-coeff-spray-series-2re-2ch}
\gamma^{(2)}_{\lambda\alpha\beta}&=&\gamma^{(2-2)}_{\lambda\alpha\beta}+\gamma^{(2-1)}_{\lambda\alpha\beta},\\
\label{sec3-cur-coeff-spray-series-2re-2ch-1}
\gamma^{(2-1)}_{\lambda\alpha\beta}&=&-2\left[A^{(2)}_{\alpha}\frac{\partial{A}^{(4)}_{\lambda}}{\partial{x}^{\beta}}
+A^{(2)}_{\beta}\frac{\partial{A}^{(4)}_{\lambda}}{\partial{x}^{\alpha}}\right]
+\frac{1}{2}\left[A^{(2)}_{\alpha\beta}\frac{\partial{A}^{(4)}}{\partial{x}^{\lambda}}
+4\left(A^{(2)}_{\alpha}\frac{\partial{A}^{(4)}_{\beta}}{\partial{x}^{\lambda}
}+A^{(2)}_{\beta}\frac{\partial{A}^{(4)}_{\alpha}}{\partial{x}^{\lambda}}\right)\right].
\end{eqnarray}
The expression of $\gamma^{(2-2)}_{\lambda\alpha\beta}$ is not displayed, which is the left part of $\gamma^{(2)}_{\lambda\alpha\beta}$ subtracted by $\gamma^{(2-1)}_{\lambda\alpha\beta}$. Similarly, we rewrite (\ref{sec3-cur-coeff-spray-series-3-2ch}) as
\begin{eqnarray}
\label{sec3-cur-coeff-spray-series-3re-2ch}
\gamma^{(3)}_{\lambda\alpha\beta}&=&\gamma^{(3-3)}_{\lambda\alpha\beta}+\gamma^{(3-2)}_{\lambda\alpha\beta}
+\gamma^{(3-1)}_{\lambda\alpha\beta},\\
\label{sec3-cur-coeff-spray-series-3re-2ch-1}
\gamma^{(3-1)}_{\lambda\alpha\beta}&=&-A^{(2)}_{\alpha}A^{(2)}_{\beta}\frac{\partial{A}^{(4)}}{\partial{x}^{\lambda}}.
\end{eqnarray}
$\gamma^{(3-3)}_{\lambda\alpha\beta}$ and $\gamma^{(3-2)}_{\lambda\alpha\beta}$ are the corresponding left parts of $\gamma^{(3)}_{\lambda\alpha\beta}$. We also need the inverse of $g_{\alpha\beta}$, and we suppose its inverse as
\begin{eqnarray}
\label{sec3-finsler-def-met-inv}
g^{\beta\nu}=A^{\beta\nu}+\frac{1}{A^{(2)}}h_{(1)}^{\beta\nu}+\frac{1}{[A^{(2)}]^{2}}h_{(2)}^{\beta\nu}+\cdots.
\end{eqnarray}
For the equation
\begin{eqnarray}
\label{sec3-finsler-def-met-inv-eq}
g_{\alpha\beta}g^{\beta\nu}=\delta_{\alpha}^{\hspace{1mm}\nu},
\end{eqnarray}
requiring the coefficients of $\frac{1}{[A^{(2)}]^{s}}$ are equivalent, we obtain
\begin{eqnarray}
\label{sec3-finsler-def-met-inv-sol-0-eq}
\frac{1}{[A^{(2)}]^{0}}:\hspace{2mm}\delta_{\alpha}^{\hspace{1mm}\nu}&=&A_{\alpha\beta}A^{\beta\nu},\\
\label{sec3-finsler-def-met-inv-sol-1-eq}
\frac{1}{[A^{(2)}]^{1}}:\hspace{4.5mm}0&=&A_{\alpha\beta}h_{(1)}^{\beta\nu}+g^{(1)}_{\alpha\beta}A^{\beta\nu},\\
\label{sec3-finsler-def-met-inv-sol-2-eq}
\frac{1}{[A^{(2)}]^{2}}:\hspace{4.5mm}0&=&A_{\alpha\beta}h_{(2)}^{\beta\nu}+g^{(1)}_{\alpha\beta}h_{(1)}^{\beta\nu}
+g^{(2)}_{\alpha\beta}A^{\beta\nu}.
\end{eqnarray}
Here only the first three equations are displayed. From the above, we know that the inverse metric can be solved term by term. The first two terms are given as
\begin{eqnarray}
\label{sec3-finsler-def-met-inv-sol-1}
h_{(1)}^{\beta\nu}&=&-6A^{\beta\rho}A^{\nu\sigma}A^{(4)}_{\rho\sigma},\\
\label{sec3-finsler-def-met-inv-sol-2}
h_{(2)}^{\beta\nu}&=&-6A^{\beta\rho}A^{(4)}_{\rho\sigma}h_{(1)}^{\sigma\nu}
-A^{\beta\rho}A^{\sigma\nu}\left[15A^{(6)}_{\rho\sigma}-A^{(2)}_{\rho\sigma}-4(A^{(2)}_{\rho}A^{(4)}_{\sigma}
+A^{(4)}_{\rho}A^{(2)}_{\sigma})\right].
\end{eqnarray}
Using the inverse metric, the 2nd class of Christoffel symbol $\gamma^{\mu}_{\alpha\beta}$ also has the same pattern with $g_{\alpha\beta}$
\begin{eqnarray}
\label{sec3-cur-coeff-spray-series-1ch}
\gamma_{\lambda\alpha\beta}=\gamma^{\mu}_{(0)\alpha\beta}+\frac{1}{A^{(2)}}\gamma^{\mu}_{(1)\alpha\beta}
+\frac{1}{[A^{(2)}]^{2}}\gamma^{\mu}_{(2)\alpha\beta}
+\frac{1}{[A^{(2)}]^{3}}\gamma^{\mu}_{(3)\alpha\beta}+\cdots,
\end{eqnarray}
with
\begin{eqnarray}
\label{sec3-cur-coeff-spray-series-0-1ch}
\gamma^{\mu}_{(0)\alpha\beta}&=&A^{\mu\lambda}\gamma^{(0)}_{\lambda\alpha\beta},\\
\label{sec3-cur-coeff-spray-series-1-1ch}
\gamma^{\mu}_{(1)\alpha\beta}&=&
A^{\mu\lambda}\gamma^{(1)}_{\lambda\alpha\beta}+h_{(1)}^{\mu\lambda}\gamma^{(0)}_{\lambda\alpha\beta},\\
\label{sec3-cur-coeff-spray-series-2-1ch}
\gamma^{\mu}_{(2)\alpha\beta}&=&
A^{\mu\lambda}\gamma^{(2)}_{\lambda\alpha\beta}+h_{(1)}^{\mu\lambda}\gamma^{(1)}_{\lambda\alpha\beta}
+h_{(2)}^{\mu\lambda}\gamma^{(0)}_{\lambda\alpha\beta}.
\end{eqnarray}
Apparently, the spray coefficients $G^{\mu}$ have the similar pattern to the metric $g_{\alpha\beta}$. Through Eq.~(\ref{sec3-cur-coeff-spray}), we obtain
\begin{eqnarray}
\label{sec3-cur-coeff-spray-series}
G^{\mu}&=&G^{\mu}_{(0)}
+\frac{1}{A^{(2)}}G^{\mu}_{(1)}
+\frac{1}{[A^{(2)}]^{2}}G^{\mu}_{(2)}+\cdots,
\end{eqnarray}
with
\begin{eqnarray}
\label{sec3-cur-coeff-spray-series-def-0}
G^{\mu}_{(0)}&=&\frac{1}{2}\gamma^{\mu}_{(0)\rho\sigma}{y}^{\rho}{y}^{\sigma},\\
\label{sec3-cur-coeff-spray-series-def-1}
G^{\mu}_{(1)}&=&\frac{1}{2}A^{\mu\lambda}\gamma^{(1)}_{\lambda\rho\sigma}{y}^{\rho}{y}^{\sigma}
+\frac{1}{2}\frac{1}{A^{(2)}}A^{\mu\lambda}\gamma^{(2-1)}_{\lambda\rho\sigma}{y}^{\rho}{y}^{\sigma}
+\frac{1}{2}\frac{1}{[A^{(2)}]^{2}}A^{\mu\lambda}\gamma^{(3-1)}_{\lambda\rho\sigma}{y}^{\rho}{y}^{\sigma}.
\end{eqnarray}
We should give some comments here. From Eq.~(\ref{sec3-cur-coeff-spray-series-2re-2ch-1}), we know that the term $\gamma^{(2-1)}_{\lambda\rho\sigma}{y}^{\rho}{y}^{\sigma}$ actually includes a factor $A^{(2)}$ and $\gamma^{(3-1)}_{\lambda\rho\sigma}{y}^{\rho}{y}^{\sigma}$ includes a factor $[A^{(2)}]^{2}$, so they both contribute to $G^{\mu}_{(1)}$ instead of $G^{\mu}_{(2)}$. So the meaning of the symbol $\gamma^{(2-1)}_{\lambda\rho\sigma}$ is that it is included in the coefficient of $\frac{1}{[A^{(2)}]^{2}}$ of $\gamma_{\lambda\rho\sigma}$ but it contributes to $G^{\mu}_{(1)}$. The meaning of the symbol $\gamma^{(3-1)}_{\lambda\rho\sigma}$ can be understood similarly. Because of these reasons,
the expressions of $G^{\mu}_{(2)}$ is quite involved, so we do not display them explicitly here.

From the foregoing expressions, we know the Ricci flag scalar in Eq.~(\ref{sec3-cur-con-flag-scalar}) also has the similar pattern to the Finsler function in Eq.~(\ref{sec3-finsler-def}), which can be written as
\begin{eqnarray}
\label{sec3-cur-ricci-series-def}
\mathcal{R}&=&\mathcal{R}^{(0)}
+\frac{1}{A^{(2)}}\mathcal{R}^{(1)}
+\frac{1}{[A^{(2)}]^{2}}\mathcal{R}^{(2)}+\cdots.
\end{eqnarray}
The manifest expressions of $\mathcal{R}^{(0)}$, $\mathcal{R}^{(1)}$ and $\mathcal{R}^{(2)}$ will be given later on. To find an equation of motion in Finsler geometry, as first thought, we propose Eq.~(\ref{sec3-cur-con-flag-scalar}) as the equation of motion. More comments on equations of motion in Finsler geometry will be given later on. For convenience, here Eq.~(\ref{sec3-cur-con-flag-scalar}) is redisplayed as
\begin{eqnarray}
\label{sec3-cur-con-flag-scalar-series}
\mathcal{R}=\mathcal{R}^{\alpha}_{\hspace{1mm}\alpha}=\Lambda(d-1)F^2.
\end{eqnarray}
Similar to the procedure in Eq.~(\ref{sec3-finsler-def-met-inv-sol-0-eq})-(\ref{sec3-finsler-def-met-inv-sol-2-eq}), we require the equivalence of coefficients of $\frac{1}{[A^{(2)}]^{s}}$ on two sides of Eq.~(\ref{sec3-cur-con-flag-scalar-series}). For the term $\frac{1}{[A^{(2)}]^{0}}$, we obtain
\begin{eqnarray}
\label{sec3-cur-con-flag-scalar-series-eq-0}
\mathcal{R}^{(0)}&=&\Lambda(d-1)A^{(2)},\\
\label{sec3-cur-con-flag-scalar-series-eq-0-def}
\mathcal{R}^{(0)}&=&2\frac{\partial {G}_{(0)}^{\alpha}}{\partial {x}^{\alpha}}-{y}^{\mu}\frac{\partial^2 {G}_{(0)}^{\alpha}}{\partial {x}^{\mu}\partial{y}^{\alpha}}+2{G}_{(0)}^{\mu}\frac{\partial^2 {G}_{(0)}^{\alpha}}{\partial {y}^{\mu}\partial{y}^{\alpha}}-\frac{\partial{G}_{(0)}^{\alpha}}{\partial {y}^{\mu}}\frac{\partial {G}_{(0)}^{\mu}}{\partial {y}^{\alpha}}.
\end{eqnarray}
For the term $\frac{1}{[A^{(2)}]^{1}}$, we obtain
\begin{eqnarray}
\label{sec3-cur-con-flag-scalar-series-eq-1}
\mathcal{R}^{(1)}&=&\Lambda(d-1)A^{(4)},\\
\label{sec3-cur-con-flag-scalar-series-eq-1-def}
\mathcal{R}^{(1)}&=&2\frac{\partial {G}_{(1)}^{\alpha}}{\partial {x}^{\alpha}}-{y}^{\mu}\frac{\partial^2 {G}_{(1)}^{\alpha}}{\partial {x}^{\mu}\partial{y}^{\alpha}}+2\left[{G}_{(0)}^{\mu}\frac{\partial^2 {G}_{(1)}^{\alpha}}{\partial {y}^{\mu}\partial{y}^{\alpha}}+{G}_{(1)}^{\mu}\frac{\partial^2 {G}_{(0)}^{\alpha}}{\partial {y}^{\mu}\partial{y}^{\alpha}}\right]-2\frac{\partial{G}_{(1)}^{\alpha}}{\partial {y}^{\mu}}\frac{\partial {G}_{(0)}^{\mu}}{\partial {y}^{\alpha}}.
\end{eqnarray}
For the term $\frac{1}{[A^{(2)}]^{2}}$, we obtain
\begin{eqnarray}
\label{sec3-cur-con-flag-scalar-series-eq-2}
\mathcal{R}^{(2)}=\Lambda(d-1)A^{(6)}.
\end{eqnarray}
The expression of $\mathcal{R}^{(2)}$ can be derived by collecting the coefficients of $\frac{1}{[A^{(2)}]^{2}}$ carefully. It does not have a simple formulation as $\mathcal{R}^{(0)}$ and $\mathcal{R}^{(1)}$. It is quite complicated as $G^{\mu}_{(2)}$, so we do not display it explicitly here.

Now we can give some analysis on the foregoing equations. The two sides of Eq.~(\ref{sec3-cur-con-flag-scalar-series-eq-0}) are polynomials of degree 2 about ${y}^{\mu}$. Their coefficients give the eqution
\begin{eqnarray}
\label{sec3-cur-con-flag-scalar-series-eq-0-compo}
\mathcal{R}^{(0)}_{\mu\nu}=\Lambda(d-1)A^{(2)}_{\mu\nu}.
\end{eqnarray}
From Eq.~(\ref{sec3-cur-con-flag-scalar-series-eq-0-def}), we know that $\mathcal{R}^{(0)}_{\mu\nu}$ is just the Ricci curvature of the spin-2 field $A_{\mu\nu}$, so Eq.~(\ref{sec3-cur-con-flag-scalar-series-eq-0-compo}) is just the equation of motion of constant curvature as in general relativity. The two sides of Eq.~(\ref{sec3-cur-con-flag-scalar-series-eq-1}) are polynomials of degree 4 about ${y}^{\mu}$. Their coefficients give the equation
\begin{eqnarray}
\label{sec3-cur-con-flag-scalar-series-eq-1-compo}
\mathcal{R}^{(1)}_{(\mu\nu\alpha\beta)}=\Lambda(d-1)A^{(4)}_{\mu\nu\alpha\beta}.
\end{eqnarray}
Here the symbol $T_{(\mu\nu\alpha\beta)}$ stands for the symmetrization of its indices. This is an equation for the spin-4 field $A_{\mu\nu\alpha\beta}$. While the two sides of Eq.~(\ref{sec3-cur-con-flag-scalar-series-eq-2}) are polynomials of degree 6 about ${y}^{\mu}$, and we have the equation
\begin{eqnarray}
\label{sec3-cur-con-flag-scalar-series-eq-2-compo}
\mathcal{R}^{(2)}_{(\mu\nu\alpha\beta\rho\sigma)}=\Lambda(d-1)A^{(6)}_{\mu\nu\alpha\beta\rho\sigma}.
\end{eqnarray}
This is an equation for the spin-6 field $A_{\mu\nu\alpha\beta\rho\sigma}$. The spin-2 field $A_{\mu\nu}$ can be solved by Eq.~(\ref{sec3-cur-con-flag-scalar-series-eq-0-compo}); the higher-spin fields $A_{\mu\nu\alpha\beta}$ and $A_{\mu\nu\alpha\beta\rho\sigma}$ can be solved by Eq.~(\ref{sec3-cur-con-flag-scalar-series-eq-1-compo}) and (\ref{sec3-cur-con-flag-scalar-series-eq-2-compo}) respectively. Therefore, through requiring the equivalence of the coefficients of $\frac{1}{[A^{(2)}]^{s}}$, Eq.~(\ref{sec3-cur-con-flag-scalar-series}) can be interpreted as a series of equations dictating higher-spin fields; while the effects of ${y}^{\mu}$ decouple from the foregoing equations completely. An important point is whether the linearization of these equations can coincide with the massless equations of motion derived by Fronsdal~\cite{Fronsdal:1978rb}. We consider the vacuum solution of zero curvature. For Eq.~(\ref{sec3-cur-con-flag-scalar-series-eq-0-compo}), we have the solution
\begin{eqnarray}
\label{sec3-cur-con-flag-scalar-series-eq-0-compo-sol}
A_{\mu\nu}=\eta_{\mu\nu},\hspace{2mm}\Lambda=0.
\end{eqnarray}
For $\Lambda=0$ and $A_{\mu\nu}=\eta_{\mu\nu}$, the linearization of Eq.~(\ref{sec3-cur-con-flag-scalar-series-eq-1-compo}) gives
\begin{eqnarray}
\label{sec3-cur-con-flag-scalar-series-eq-1-compo-linear}
\partial^{\rho}\partial_{\rho}A_{\mu\nu\alpha\beta}-4\partial_{(\mu}\partial^{\rho}A_{\rho\nu\alpha\beta)}
+6\partial_{(\mu}\partial_{\nu}\eta^{\rho\sigma}A_{\rho\sigma\alpha\beta)}=0,
\end{eqnarray}
where the symbol $T_{(\mu\nu\alpha\beta)}$ stands for the symmetrization of the un-contracted indices and the indices are contracted by $\eta^{\rho\sigma}$. Eq.~(\ref{sec3-cur-con-flag-scalar-series-eq-1-compo-linear}) is just the conventional Fronsdal equation for the spin-4 field~\cite{Fronsdal:1978rb}.

Different from the scalar type of equations in Eq.~(\ref{sec3-cur-con-flag-scalar-series}), the Einstein type of equations of motion in Finsler geometry have been proposed in series of papers~\cite{Asanov:1981Fo,Vacaru:2008qs,miron1994,Chang:2008yv,Pfeifer:2011xi}. Those equations are tensor types with 2 indices. They generally have stronger constraints than Eq.~(\ref{sec3-cur-con-flag-scalar-series}), but do not have a simple interpretation as equations of motion for higher-spin fields as we displayed above.

\subsection{Nonlinear Extensions}\label{sec:3.3}

From section \ref{sec:3.2}, we learned that Eq.~(\ref{sec3-cur-con-flag-scalar-series-eq-0}) is just the equation of constant curvature for the spin-2 field. It does not acquire any corrections from higher-spin fields. Eq.~(\ref{sec3-cur-con-flag-scalar-series-eq-1}) includes the spin-2 field and the spin-4 field, and it is linear about the spin-4 field. Eq.~(\ref{sec3-cur-con-flag-scalar-series-eq-2}) includes the spin-2 field, the spin-4 field and the spin-6 field, and it is also linear about the spin-6 field. So what we obtained from Eq.~(\ref{sec3-cur-con-flag-scalar-series}) are free equations of motion for higher-spin fields. In order to embody the interactions of higher-spin fields, we need some non-linear extensions of Eq.~(\ref{sec3-cur-con-flag-scalar-series}). For example, noticing that $\mathcal{R}$ is a homogeneous function of degree 2 of ${y}^{\mu}$, we can supplement Eq.~(\ref{sec3-cur-con-flag-scalar-series}) with a non-linear term
\begin{eqnarray}
\label{sec3-cur-con-flag-scalar-series-eq-non-linear-2}
\Delta_{2}=(g^{\alpha\beta}g^{\mu\nu}+g^{\alpha\mu}g^{\beta\nu}+g^{\alpha\nu}g^{\mu\beta})
\frac{\partial^4[F^2\mathcal{R}^2]}{\partial{y}^{\alpha}\partial{y}^{\beta}\partial{y}^{\mu}\partial{y}^{\nu}}.
\end{eqnarray}
This term is also a homogeneous function of degree 2 of ${y}^{\mu}$. It supplies higher curvature corrections for Eq.~(\ref{sec3-cur-con-flag-scalar-series-eq-0}), and also supplies quadratic non-linear corrections with higher derivatives for Eq.~(\ref{sec3-cur-con-flag-scalar-series-eq-1}). Apparently, lots of terms similar to Eq.~(\ref{sec3-cur-con-flag-scalar-series-eq-non-linear-2}) can be constructed. On the other hand, from the efforts to construct cubic interaction vertices for higher-spin fields, we learned that the interactions for higher-spin fields involve higher derivatives and also restricted by no-go theorems~\cite{Bekaert:2010hw}. Therefore, it keeps as a challenge to find a general guiding principle to organize these non-linear corrections as Eq.~(\ref{sec3-cur-con-flag-scalar-series-eq-non-linear-2}) systematically and consistently.

\section{Discussions and Conclusions}\label{sec:4}

Unlike the concise formulation of the spin-1 gauge fields and the spin-2 gravity, the formulations of higher-spin fields are more involved, as shown by the nonlinear equations of motion constructed by Vasiliev based on the frame-like fields~\cite{Vasiliev:1990en,Vasiliev:2003ev}. A geometrical formulation of higher-spin fields appears to be more difficult to be constructed, even at the level of equations of motion. The analysis on objects of constant curvature show that Finsler geometry could be useful tools to explore geometrical formulations of higher-spin fields, as is supported by the observation that Fronsdal's equations can be derived from the equation of constant curvature in Finsler geometry in section \ref{sec:3.2}. However, in order to construct a complete formulation of higher-spin fields based on Finsler geometry, guiding principles organizing nonlinear corrections as considered in section \ref{sec:3.3} are required. In this regard, the method used in~\cite{Joung:2012hz} may be relevant. Besides discussions on equations of motion in section \ref{sec:3}, an consistent action which can generate those equations is required. Several actions has been proposed~\cite{Asanov:1981Fo,Asanov:1982Fo,Asanov:1983Fo,Pfeifer:2011xi}, albeit actions solely based on higher-spin fields seem difficult to be constructed. The difficulty in this situation is similar to that of constructing actions for Vasiliev's equations~\cite{Boulanger:2011dd,Doroud:2011xs,Vasiliev:2011zza}.

\acknowledgments

This work is supported by Project Basal under Contract No.~FB0821.

\bibliographystyle{utphys}

\bibliography{HighSpinFinslerRef}

\end{document}